\documentstyle[12pt]{article}
\newcommand{\JHEP}{J. High Energy Phys. }

\newcommand{\NP}{Nucl. Phys. }

\newcommand{\PL}{Phys. Lett. }
\newcommand{\PTP}{Prog. Theor. Phys. }
\newcommand{\RMP}{Rev. Mod. Phys. }

\begin{document}
\baselineskip=20pt

\pagenumbering{arabic}

\vspace{1.0cm}
\begin{flushright}
LU-ITP 2004/001
\end{flushright}

\begin{center}
{\Large\sf On double gauging of $U(1)$ symmetry on noncommutative space}\\[10pt]
\vspace{.5 cm}

{Yi Liao, Klaus Sibold}
\vspace{1.0ex}

{\small Institut f\"ur Theoretische Physik, Universit\"at Leipzig,
\\
Augustusplatz 10/11, D-04109 Leipzig, Germany\\}

\vspace{2.0ex}

{\bf Abstract}

\end{center}

We point out that a field $\varphi$ charged under a global $U(1)$ symmetry
generally allows for a starred localized extension with the transformation rule,
$\varphi\to U_L\star\varphi\star U_R^{-1}$. This results in a double gauging of
the global $U(1)$ symmetry on noncommutative space. We interpret the gauge theory
so obtained in terms of the gauge fields that in the commutative limit appear
naturally and are respectively the gauge field responsible for the charge and
a decoupled vector field. The interactions are shown to be very different from
those obtained by assigning a transformation rule of $\varphi\to U\star\varphi$
or $\varphi\star U^{-1}$.

\begin{flushleft}


\end{flushleft}

\newpage

Charge is a global property of fields. Its conservation implies a global $U(1)$
symmetry. When the symmetry is gauged on ordinary spacetime, the charge of a field
specifies how the field transforms locally together with the gauge field. This
gauging procedure may be generalized to field theory on noncommutative (NC)
spacetime in the approach of the Moyal-Weyl correspondence $\cite{review}$.
For a $U(1)$ gauge field transforming as
\begin{equation}
\begin{array}{rcl}
A_{\mu}(x)&\to&\displaystyle U(x)\star A_{\mu}(x)\star U^{-1}(x)
-\frac{i}{e}U(x)\star\partial_{\mu}U^{-1}(x),
\end{array}
\label{a}
\end{equation}
the following types of matter field transformation rules have been studied
$\cite{hayakawa}$,
\begin{equation}
\begin{array}{rcl}
\psi_+(x)&\to&U(x)\star\psi_+(x),\\
\psi_-(x)&\to&\psi_-(x)\star U^{-1}(x),\\
\psi_0(x)&\to&U(x)\star\psi_0(x)\star U^{-1}(x),
\end{array}
\label{old}
\end{equation}
besides the trivial identity representation. Here the Moyal star product is
defined as
\begin{equation}
\begin{array}{rcl}
(f_1\star f_2)(x)&=&\displaystyle\left[\exp\left(\frac{i}{2}
\theta^{\mu\nu}\partial^x_{\mu}\partial^y_{\nu}\right)
f_1(x)f_2(y)\right]_{y=x},
\end{array}
\end{equation}
with $\theta_{\mu\nu}$ being the parameter characterizing the NC spacetime,
and $U(x)$ is the starred exponential,
\begin{equation}
\displaystyle
U(x)=\exp_{\star}[ie\alpha(x)]
=1+ie\alpha(x)+\frac{(ie)^2}{2!}\alpha\star\alpha(x)+\cdots.
\end{equation}
The restrictions on transformation rules and also on gauge groups originate
essentially from the closure requirement of group multiplication and the
noncommutativity of the star product between c-number functions
$\cite{terashima,armoni}$. This poses an obstacle in realistic model building
$\cite{chaichian}$.

In this work we would like to point out a new transformation rule for
matter fields that is more general than the ones listed in eqn. $(\ref{old})$,
\begin{equation}
\begin{array}{rcl}
\varphi(x)&\to&U_L(x)\star\varphi(x)\star U_R^{-1}(x),
\end{array}
\label{new}
\end{equation}
where $U_{L,R}(x)$ are two independent starred exponentials. This generalization
is based on the following observation. The fact that the $\varphi$ field carries
a conserved, additive charge is strong enough to fix uniquely its local
transformation rule on ordinary spacetime but not so on NC spacetime where the
commutative point-wise multiplication is replaced by the noncommutative star product.
The order of factors becomes relevant as we already saw in the transformation rules
for the above $\psi$ fields. On the other hand, as far as charge is
concerned, the requirement that must be met is the global transformation rule of
the charged field for which there is no difference between the point-wise and
star product. This is indeed the case for the rule in eqn. $(\ref{new})$: for
constant $U_{L,R}$, only the combination $U=U_LU_R^{-1}$ is relevant.
Furthermore, multiplying more factors like $U_{L,R}(x)$ from the left or right
in eqn. $(\ref{new})$ is ambiguous because the order of these factors, while
relevant due to the star, is not a well-defined concept $\cite{chaichian}$.
This makes eqn. $(\ref{new})$ the most general transformation that can be assumed
for a charged field.

The implementation of the transformation rule $(\ref{new})$ necessarily demands
two $U(1)$ gauge bosons. The NC $U(1)\times U(1)$ gauge theory has been studied
previously $\cite{liao}$ by assigning to a scalar field two independent charges
corresponding to the two factors of $U(1)$, which are then spontaneously broken.
We stress that our point of view is quite different in this work. We
assume for a field $\varphi$ only one charge instead of two independent ones
while introducing two sets of gauge bosons. This is not as surprising as it
seems to be at first sight since it appears in some sense already for the neutral
field $\psi_0$: we can introduce a gauge field for a matter field with no charge
at all. This possibility is offered by the spacetime structure as opposed to
field theory on ordinary spacetime in which an uncharged field cannot be engaged
in gauge interactions. To clarify our point and show its consequences, we take the
$\varphi^4$ theory as an example, but generalization to fermion fields is
straightforward.

Consider the complex $\varphi^4$ theory on NC spacetime whose Lagrangian density is
\begin{equation}
\begin{array}{rcl}
{\cal L}&=&\displaystyle
\partial_{\mu}\varphi\star\partial^{\mu}\varphi^{\dagger}
-m^2\varphi\star\varphi^{\dagger}
-\frac{\lambda}{2}\varphi\star\varphi^{\dagger}\star\varphi\star\varphi^{\dagger}.
\end{array}
\label{phi4}
\end{equation}
Its equations of motion are derived by the variational principle generalized to
NC spacetime (see Ref. $\cite{micu}$ for example) as
\begin{equation}
\begin{array}{rcl}
\delta S&=&\displaystyle
\int d^4x~\delta{\cal L}\\
&=&\displaystyle
\int d^4x\left\{-\delta\varphi\star(\partial^2+m^2)\varphi^{\dagger}
-(\partial^2+m^2)\varphi\star\delta\varphi^{\dagger}\right.\\
&&\displaystyle
-\frac{\lambda}{2}\left(
\delta\varphi\star\varphi^{\dagger}\star\varphi\star\varphi^{\dagger}
+\varphi\star\varphi^{\dagger}\star\delta\varphi\star\varphi^{\dagger}\right.\\
&&\displaystyle\left.\left.
~~~~+\varphi\star\delta\varphi^{\dagger}\star\varphi\star\varphi^{\dagger}
+\varphi\star\varphi^{\dagger}\star\varphi\star\delta\varphi^{\dagger}\right)
\right\}.
\end{array}
\end{equation}
Using the cyclicity property of integrals of star products, the interaction terms
can be combined and one star can be ignored in each term,
\begin{equation}
\begin{array}{rcl}
\delta S&=&\displaystyle
-\int d^4x\left\{\delta\varphi\left((\partial^2+m^2)\varphi^{\dagger}
+\lambda\varphi^{\dagger}\star\varphi\star\varphi^{\dagger}\right)\right.\\
&&\displaystyle
~~~~~~~~\left.
+\left((\partial^2+m^2)\varphi
+\lambda\varphi\star\varphi^{\dagger}\star\varphi\right)
\delta\varphi^{\dagger}\right\}.
\end{array}
\end{equation}
The equations of motion are then,
\begin{equation}
(\partial^2+m^2)\varphi=-\lambda\varphi\star\varphi^{\dagger}\star\varphi
\label{eom}
\end{equation}
and its Hermitian conjugate.

The $\varphi^4$ theory in eqn. $(\ref{phi4})$ has a global $U(1)$ symmetry. Now
we construct its conserved currents by the Noether procedure. For this purpose,
we make the global transformation star localized. As discussed above, this is
not unique. Consider first an infinitesimal transformaion from the right with
$U_R(x)=\exp_{\star}[-i\alpha_R(x)]=1-i\alpha_R(x)+\cdots$. Note that only the
kinetic term contributes to the variation of the action. Using again the cyclicity
property and ignoring one star in each term, we have
\begin{equation}
\begin{array}{rcl}
\delta S&=&\displaystyle\int d^4x~\partial^{\mu}\alpha_R
~i(\partial_{\mu}\varphi^{\dagger}\star\varphi
-\varphi^{\dagger}\star\partial_{\mu}\varphi)
+O(\alpha_R^2),
\end{array}
\end{equation}
which determines the right current to be
\begin{equation}
\begin{array}{rcl}
J_{\mu}^R&=&\displaystyle
-i(\partial_{\mu}\varphi^{\dagger}\star\varphi
-\varphi^{\dagger}\star\partial_{\mu}\varphi).
\end{array}
\end{equation}
The left current is similarly obtained from the left transformation,
\begin{equation}
\begin{array}{rcl}
J_{\mu}^L&=&\displaystyle
+i(\partial_{\mu}\varphi\star\varphi^{\dagger}
-\varphi\star\partial_{\mu}\varphi^{\dagger}).
\end{array}
\end{equation}
Both currents are conserved using the equations of motion in eqn. $(\ref{eom})$
and its Hermitian conjugate.

The left and right currents are related by
$\varphi\leftrightarrow\varphi^{\dagger}$ up to arbitrary normalization factors.
Actually they share the same charge. It would then be tempting to introduce two
other currents,
\begin{equation}
\begin{array}{rcl}
J^{\pm}_{\mu}&=&\displaystyle\frac{1}{2}\left(J^R_{\mu}\pm J^L_{\mu}\right),
\end{array}
\end{equation}
which would correspond to the following two independent $U(1)$ transformations,
\begin{equation}
\begin{array}{rcl}
\varphi(x)&\to&U_1(x)\star\varphi(x)\star U_1(x),\\
\varphi(x)&\to&U_2(x)\star\varphi(x)\star U_2^{-1}(x),
\end{array}
\label{wrong}
\end{equation}
under which $\varphi$ is charged and neutral respectively. The problem is that,
when put together, it is ambiguous how to order the two transformations as
mentioned above. Further, it is impossible to associate the first transformation
with a real gauge boson which should transform as in eqn. $(\ref{a})$.
Thus, to gauge the symmetry consistently we should implement the transformation
rule in eqn. $(\ref{new})$ instead of the one in eqn. $(\ref{wrong})$.

It is now standard to write down the Lagrangian for the doubly gauged $\varphi^4$
theory. The two gauge bosons are denoted as $L_{\mu}$ and $R_{\mu}$ with gauge
couplings $g_{L,R}$. Together with the $\varphi$ transformation, we have
\begin{equation}
\begin{array}{rcl}
L_{\mu}&\to&\displaystyle
U_L(x)\star L_{\mu}\star U_L^{-1}(x)
+\frac{i}{g_L}U_L(x)\star\partial_{\mu}U_L^{-1}(x),\\
R_{\mu}&\to&\displaystyle
U_R(x)\star R_{\mu}\star U_R^{-1}(x)
+\frac{i}{g_R}U_R(x)\star\partial_{\mu}U_R^{-1}(x).
\end{array}
\label{vector}
\end{equation}
The covariant derivative and field tensors are
\begin{equation}
\begin{array}{rcl}
D_{\mu}\varphi&=&\displaystyle
\partial_{\mu}\varphi-ig_LL_{\mu}\star\varphi+ig_R\varphi\star R_{\mu},\\
L_{\mu\nu}&=&\displaystyle
\partial_{\mu}L_{\nu}-\partial_{\nu}L_{\mu}-ig_L[L_{\mu},L_{\nu}]_{\star},\\
R_{\mu\nu}&=&\displaystyle
\partial_{\mu}R_{\nu}-\partial_{\nu}R_{\mu}-ig_R[R_{\mu},R_{\nu}]_{\star},
\end{array}
\end{equation}
with $[f_1,f_2]_{\star}=f_1\star f_2-f_2\star f_1$. Then,
\begin{equation}
\begin{array}{rcl}
{\cal L}&=&\displaystyle{\cal L}_{\varphi}+{\cal L}_L+{\cal L}_R,\\
{\cal L}_{\varphi}&=&\displaystyle
D_{\mu}\varphi\star (D^{\mu}\varphi)^{\dagger}-m^2\varphi\star\varphi^{\dagger}
-\frac{\lambda}{2}\varphi\star\varphi^{\dagger}\star\varphi
\star\varphi^{\dagger},\\
{\cal L}_L&=&\displaystyle
-\frac{1}{4}L_{\mu\nu}\star L^{\mu\nu},\\
{\cal L}_R&=&\displaystyle
-\frac{1}{4}R_{\mu\nu}\star R^{\mu\nu}.
\end{array}
\end{equation}
The action $\displaystyle S=\int d^4x {\cal L}$ is gauge invariant but not the
Lagrangian density itself,
\begin{equation}
\begin{array}{lll}
{\cal L}_{\varphi}\to U_L\star{\cal L}_{\varphi}\star U_L^{-1},
&{\cal L}_L\to U_L\star{\cal L}_L\star U_L^{-1},
&{\cal L}_R\to U_R\star{\cal L}_R\star U_R^{-1}.
\end{array}
\end{equation}
The apparent asymmetry for ${\cal L}_{\varphi}$ in the left and right
transformations arises because we have arbitrarily ordered the $\varphi$ field
in such a way that $\varphi$ comes first in each term of ${\cal L}_{\varphi}$.
This is immaterial of course.

Now comes the question of how to interpret the above gauge theory. As we
emphasized above, the left and right currents correspond to the same charge.
In this sense the two gauge bosons $L_{\mu}$ and $R_{\mu}$ stand on the same
footing. On the other hand, to a single {\it electric} charge we certainly
expect to associate a single gauge field that can be called {\it the}
electromagnetic field. To solve this dilemma, we note that what determines the
electric charge of the $\varphi$ field is its global transformation, for which
there is no difference between ordinary and starred products. This implies that
the physical fields can be correctly identified by going to the commutative
limit of $\theta_{\mu\nu}\to 0$ in the Lagrangian density. In this limit,
we have
\begin{equation}
\begin{array}{rcl}
D_{\mu}\varphi&\to&\displaystyle
\partial_{\mu}\varphi-ieA_{\mu}\varphi,
\end{array}
\end{equation}
where $A_{\mu}$ can be identified unambiguously with the electromagnetic field
with the electric coupling $e=\sqrt{g_L^2+g_R^2}$. The orthogonal $B_{\mu}$
field drops from the covariant derivative and is thus not related with the
electric charge. Both fields are linear combinations of the original gauge
bosons,
\begin{equation}
\displaystyle
\left(\begin{array}{c}A_{\mu}\\B_{\mu}
      \end{array}
\right)=
\left(\begin{array}{cc}c&-s\\s&c
      \end{array}
\right)
\left(\begin{array}{c}L_{\mu}\\R_{\mu}
      \end{array}
\right),
\label{rotation}
\end{equation}
with $c=g_L/e,~s=g_R/e$. Note that in the above limit the gauge terms
simplify into the free kinetic terms quadratic in $L_{\mu},~R_{\mu}$ and thus
also in $A_{\mu},~B_{\mu}$, hence the $B_{\mu}$ becomes a decoupled, harmless
free field. This also means in passing that we will need a genuine NC vertex
to normalize the other coupling parameter $c$ or $s$ which does not show up
in the commutative limit.

Once the physical fields are identified, we should reexpress in terms of them
the Lagrangian on NC spacetime. The relevant pieces are,
\begin{equation}
\begin{array}{rcl}
D_{\mu}\varphi&=&\displaystyle
\partial_{\mu}\varphi-ie(c^2A_{\mu}\star\varphi+s^2\varphi\star A_{\mu})
+iecs[\varphi,B_{\mu}]_{\star},\\
L_{\mu\nu}&=&\displaystyle
c(\partial_{\mu}A_{\nu}-\partial_{\nu}A_{\mu}-iec^2[A_{\mu},A_{\nu}]_{\star})\\
&+&\displaystyle
s(\partial_{\mu}B_{\nu}-\partial_{\nu}B_{\mu}-iecs[B_{\mu},B_{\nu}]_{\star})\\
&-&\displaystyle
iec^2s([A_{\mu},B_{\nu}]_{\star}+[B_{\mu},A_{\nu}]_{\star}),\\
R_{\mu\nu}&=&\displaystyle
c(\partial_{\mu}B_{\nu}-\partial_{\nu}B_{\mu}-iecs[B_{\mu},B_{\nu}]_{\star})\\
&-&\displaystyle
s(\partial_{\mu}A_{\nu}-\partial_{\nu}A_{\mu}+ies^2[A_{\mu},A_{\nu}]_{\star})\\
&+&\displaystyle
iecs^2([A_{\mu},B_{\nu}]_{\star}+[B_{\mu},A_{\nu}]_{\star}).
\end{array}
\end{equation}
We notice some features in the Lagrangian formed from the above pieces. With
$c=0,~s=1$ or $c=1,~s=0$ we recover a theory that would have been derived by
following the transformation rules for $\psi_{\pm}$, plus a vector field
$B_{\mu}$ that is completely free. Except for these two special cases, the theory
is very different. First, the photon-matter interaction terms in $D_{\mu}\varphi$
appear in both orders with generally different weights, which makes the
interactions even richer than the usual ones. Second, when omitting the mixing
terms in the gauge terms, we find that the $B$ terms are in a canonical form as
would be obtained by gauging a $U(1)$ directly. But there is no way for the $A$
terms to become canonical. This is consistent with our discussions following
eqn. $(\ref{wrong})$.
That is to say, the symmetry properties are simplest in terms of the
$L_{\mu},R_{\mu}$ fields which the physical interpretation can be better
expressed in terms of the $A_{\mu},B_{\mu}$ fields.

To complete the gauge theory, we should add the gauge fixing terms and the
corresponding ghost terms. They can be obtained by generalizing directly the
formalism of Kugo and Ojima on ordinary spacetime $\cite{kugo}$ which utilizes
Hermitian ghost and anti-ghost fields,
\begin{equation}
\begin{array}{rcl}
{\cal L}_{\rm g.f.}+{\cal L}_{\rm ghost}&=&\displaystyle
i{\sf s}\left(-\frac{1}{2}(\xi_L\bar{c}_L\star h_L+\xi_R\bar{c}_R\star h_R)
+(\partial^{\mu}\bar{c}_L)\star L_{\mu}+(\partial^{\mu}\bar{c}_R)\star R_{\mu}
\right)\\
&=&\displaystyle
-(\partial^{\mu}h_L\star L_{\mu}+\partial^{\mu}h_R\star R_{\mu})
+\frac{1}{2}\left(\xi_L h_L\star h_L+\xi_R h_R\star h_R\right)\\
&&\displaystyle
-i(\partial^{\mu}\bar{c}_L)\star D_{\mu}c_L
-i(\partial^{\mu}\bar{c}_R)\star D_{\mu}c_R,
\end{array}
\end{equation}
where $h_{L,R}$ are auxiliary fields, $\xi_{L,R}$ are gauge parameters, and
$D_{\mu}c_L=\partial_{\mu}c_L+ig_L[c_L,L_{\mu}]_{\star}$,
$D_{\mu}c_R=\partial_{\mu}c_R+ig_R[c_R,R_{\mu}]_{\star}$.
The nilpotent BRS variations are ($j=L,R$),
\begin{equation}
\begin{array}{rcl}
{\sf s}\varphi&=&\displaystyle
ig_Lc_L\star\varphi-ig_R\varphi\star c_R,\\
{\sf s}\varphi^{\dagger}&=&\displaystyle
ig_Rc_R\star\varphi^{\dagger}-ig_L\varphi^{\dagger}\star c_L,\\
{\sf s}L_{\mu}&=&\displaystyle
D_{\mu}c_L,\\
{\sf s}R_{\mu}&=&\displaystyle
D_{\mu}c_R,\\
{\sf s}c_j&=&ig_jc_j\star c_j,\\
{\sf s}\bar{c}_j&=&ih_j,\\
{\sf s}h_j&=&0.
\end{array}
\label{brs}
\end{equation}
The BRS currents and charge can also be constructed. We found that the BRS
currents generally contain terms which are not in a closed form in terms of
star product but are cumbersome series in the NC parameter. However, when time
is not involved in the noncommutativity, all these terms do not contribute to
the conserved BRS charge,
\begin{equation}
\begin{array}{rcl}
Q_{\rm BRS}&=&\displaystyle
\int d^3{\bf x}\sum_{j=L,R}\left(
-\dot{h}_j\star c_j+h_j\star D_0c_j+g_j\dot{\bar{c}}_j\star c_j\star c_j
\right)(x),
\end{array}
\end{equation}
where the dot stands for the time derivative. Using the canonical equal-time
(anti-) commutation relations we have verified that the above charge indeed
generates the BRS variations shown in eqn. $(\ref{brs})$. The physical Hilbert
space is then characterized by its annihilation by the BRS charge.

In summary, we have proposed that a matter field charged under a single $U(1)$
allows for a starred local transformation rule on NC space that acts from the
left and right independently. This makes possible a double gauging of one global
$U(1)$ symmetry on NC space. We suggested how the resulting theory should be
interpreted in terms of physical degrees of freedom that are identified using
the global property of charge: one gauge boson interacts with the charge while
the other interacts only noncommutatively and thus decouples on ordinary
space. We have also shown that the interactions in this theory have a richer
structure than those obtained by a direct gauging of the charge $U(1)$ symmetry.

\end{document}